\documentclass[a4]{elsart5p}

\usepackage{amssymb,amsmath} 
\usepackage{bm,hyphenat,xspace} %,draftcopy}
\usepackage{graphicx,epsfig}
\usepackage{tabularx}
\newcommand{\email}[1]{\ead{#1}}

\newcommand {\mcu}{\mathcal{U}}
\newcommand {\mcl}{\mathcal{L}}
\newcommand {\mcj}{\mathcal{J}}
\newcommand {\mct}{T}

\begin{document}
%\linenumbers

\begin{frontmatter}

\title {Recombination in the universal four-fermion system}

\author{A.~Deltuva}
\email{arnoldas.deltuva@tfai.vu.lt}

%\affiliation{
\address{
Institute of Theoretical Physics and Astronomy, 
Vilnius University, Saul\.etekio al. 3, LT-10257 Vilnius, Lithuania}
%Centro de F\'{\i}sica Nuclear da Universidade de Lisboa, 
%P-1649-003 Lisboa, Portugal }

%\received{April 11, 2018}

%\pacs{24.10.-i, 21.45.-v, 25.45.Hi, 25.40.Hs}

\begin{abstract}
  In the systems of spin $\frac12$ fermions with resonant $S$-wave interactions
  supporting only weakly bound  dimers the antisymmetry forbids recombination
  of three (or more) fermions at zero energy.
  However, the fermion-fermion-dimer recombination is only partially suppressed.
  It is studied in the framework of momentum-space integral equations for 
  the four-particle transition operators. In the vicinity of the unitary limit
  the fermion-fermion-dimer recombination rate,  rescaled to build dimensionless quantity,
  is found to be  linear in the effective range parameter,
  enabling a simple and accurate parametrization as well as  evaluation of finite-range effects
  for any potential model. This feature makes the present results very useful
 in benchmarking different methods for three-cluster breakup and recombination 
calculations in four-particle systems.
The  interplay of the three-fermion and fermion-fermion-dimer recombination
processes and their consequences for ultracold mixtures of fermions and dimers
is discussed.
\end{abstract}


\end{frontmatter}
% \maketitle

% -----------------

\section{Introduction \label{sec:intro}}

Universality in few-body systems with resonant $S$-wave interactions, characterized by a large scattering length, 
has been investigated in numerous works, where the considered systems range from cold atoms and molecules to
nuclear and particle physics \cite{braaten:rev,hammer:10a,naidon:rev,greene:rev,kievsky:rev21}.
The qualitative behavior of those physical systems depends critically on the permutation symmetry they obey,
i.e., the bosonic/fermionic nature.
While bosons (or distinguishable particles) exhibit a rich spectrum of few-body bound states,
the most famous example being the three-body Efimov effect \cite{efimov:plb},
identical spin $\frac12$ fermions (unpolarized, i.e., in both spin states)
can form only one shallow bound state, 
a ${}^1S_0$ dimer with the binding energy $b_d \approx \hbar^2/ma^2$, where $m$ is the fermion mass
and $a$ is the two-fermion scattering length.
Although there are no three- and four-fermion bound 
states\footnote{This work assumes the absence of  non universal deeply bound two-body states, 
implying also the absence of the associated many-body states},
 the three- and four-body physics is important for the properties
of cold dilute atomic and molecular gases that are determined by the parameters
of low-energy  collisions
\cite{petrov:04a,PhysRevA.77.043619,PhysRevA.79.030501,STRINATI20181}.
For example, the fermionic dimer-dimer scattering  at low energy, well below the breakup threshold, 
has been investigated in a number of works 
\cite{petrov:04a,PhysRevA.77.043619,PhysRevA.79.030501,bulgac:03a,levinsen:11a,elhatisari:17a, deltuva:17d},
achieving a good agreement between the different methods for the dimer-dimer scattering length and for the 
effective range, except for the lattice effective field theory approach \cite{elhatisari:17a} for the latter observable.
The present work focuses on a different aspect of the four-fermion system near unitarity, namely,
the production of fermionic dimers in few-body collisions.
In terms of initial state complexity, the most straightforward path is the collision of two fermions in different
spin states; however, the formation of a dimer necessitates an additional interaction with an external field,
e.g., with the electromagnetic field where a photon is radiated thereby ensuring the energy and momentum conservation.
If this interaction is suppressed for some reason and the only forces retained are those acting between the
particles themselves, the creation of a dimer may proceed via recombination process involving at least three
colliding particles or clusters.
However, the antisymmetry requirement suppresses the recombination of three spin $\frac12$ fermions in the ultracold limit
since the zero kinetic energy state of three (or more) fermions can not be fully antisymmetric;
the dominant recombination channel has total orbital momentum $\mcl=1$ where the recombination rate 
vanishes at the threshold since it scales with the energy \cite{esry:01a}.
 The above antisymmetry restrictions do not apply and zero-energy recombination
is possible in the collision of two fermions with a third distinct particle. However, such a system loses its
fermionic character becoming effectively a system of three distinguishable particles; it is therefore out of the 
interest for the present study, except for a very special choice of the third particle, namely a dimer consisting
of the same kind of fermions as the first two. This choice preserves the purely fermionic character of the system
but  implies right away a four-body problem. 
The zero kinetic energy state in the two-fermion plus dimer system is not precluded
such that the recombination rate does not vanish in the ultracold limit. However, one may
expect partial suppression due to the antisymmetry of the system, since the wave function must be antisymmetric also
under the exchange of the free fermions with those building the dimer, a condition that puts restrictions
on the three-cluster system as well. It is therefore interesting to study the  fermion-fermion-dimer recombination
as a four-fermion problem, to evaluate its importance relative to other possible channels  and consequences
to ultracold mixtures of fermionic atoms and diatomic molecules. Furthermore, the results in the unitary limit 
are expected to be universal, i.e., independent of the short-range interaction details, and therefore perfectly
suitable for the future benchmark calculations of the three-cluster recombination or its time-reverse 
dimer-dimer breakup reaction in the four-fermion system.

Section II outlines essential details of the four-fermion collision
 calculations, and   Sec. III presents the obtained results for the recombination process.
 The summary is given in Sec. IV.

\section{Theory \label{sec:eq}}

Elastic scattering of two fermionic dimers was considered
in Ref.~\cite{deltuva:17d} providing most accurate results for the
dimer-dimer effective range and phase shifts at finite energy below the
breakup threshold. That work employed rigorous Faddeev-Yakubovsky theory
\cite{yakubovsky:67} for the four-particle scattering in the integral form
proposed by Alt, Grassberger, and Sandhas (AGS) \cite{grassberger:67}. 
The AGS framework describes the four-particle scattering problem in terms
of subsystem transition operators, namely, the two-particle, i.e., 2+1+1, transition matrix
\begin{equation} \label{eq:t}
  t= v + v G_0 t,
\end{equation}
the three-particle, i.e., 3+1 with $\alpha=1$, and two-pair, i.e., 2+2 with $\alpha=2$,
transition operators
\begin{equation} \label{eq:U3}
U_{\alpha} =  P_\alpha G_0^{-1} + P_\alpha  t G_0  U_{\alpha}.
\end{equation}
The resulting four-particle transition operators obey the  symmetrized AGS equations; the subset relevant
for the present consideration of reactions involving two dimers reads
\begin{subequations} \label{eq:U}
\begin{align}  
\mcu_{12}  = {}&  (G_0  t  G_0)^{-1}  
 - P_{34}  U_1 G_0  t G_0  \mcu_{12} + U_2 G_0  t G_0  \mcu_{22} , 
\label{eq:U12} \\  
\mcu_{22}  = {}& (1 - P_{34}) U_1 G_0  t  G_0  \mcu_{12} . \label{eq:U22}
\end{align}
\end{subequations}
In the above equations the subscripts $\alpha= 1$ (2) label the clustering of the 3+1 (2+2) type,
$G_0 = (E+i0-H_0)^{-1}$ is the free  four-particle resolvent with the 
energy $E$  and the kinetic energy operator $H_0$,
both in the center-of-mass (c.m.) frame, $v$ is the two-particle potential,
$P_{ab}$ is the permutation operator of particles $a$ and $b$,
while $P_\alpha$ are  combinations of  permutation operators \cite{deltuva:17d} that,
together with a proper choice of basis states, ensure the desired antisymmetry
of the four-fermion system as explained  in  Ref.~\cite{deltuva:17d} and references therein.

The most efficient way to solve the  AGS equations (\ref{eq:U})
with short-range forces is by employing
the partial-wave decomposition in the momentum-space, leading to 
a system of integral equations  with three continuous variables.
Those are the  magnitudes of the 
Jacobi momenta  $k_x$ , $k_y$, and $k_z$ describing the relative motion between particles and/or 
clusters \cite{deltuva:12a}. The  corresponding  orbital angular momenta  $l_x$, $l_y$, and $l_z$
and fermion spins  $s_i = \frac12$ are coupled to build eigenstates of 
the  total angular momentum  $\mathcal{J}$ and its projection  $\mathcal{M}$.
Different coupling schemes are used for the 3+1 and 2+2 clustering,
$ | k_x \, k_y \, k_z \rangle_1 \otimes
|\{l_z [(l_y \{[l_x (s_1s_2)s_x]j_x \, s_3\}S_y ) J_y s_4 ] S_z \} \,\mathcal{JM} \rangle_1$ 
and 
$|k_x \, k_y \, k_z \rangle_2 \otimes
|(l_z  \{ [l_x (s_1s_2)s_x]j_x\, [l_y (s_3 s_4)s_y] j_y \} S_z)\mathcal{ J M} \rangle_2 $,
respectively, where the remaining quantum numbers such as $j_x$ etc.  are angular momenta 
of the intermediate subsystems as explained in Ref.~\cite{deltuva:17d}.
The antisymmetry condition restricts
$l_x + s_x$ (and $l_y + s_y$ for the $2+2$ configuration) to even values;
$l_x=s_x=j_x=0$ for the $S$-wave dimer considered in the present work.

An important aspect of the solution is the treatment of integrable kernel singularities
in Eqs.~(\ref{eq:U}) arising in $t$ and $U_2$ due to the one- and two-dimer poles, respectively.
Their treatment using the complex-energy method with special integration weights is taken over
from Ref.~\cite{deltuva:12c}.

The amplitude for the elastic dimer-dimer scattering, its relation to the phase shift, and effective-range
expansion parameters can be found in Ref.~\cite{deltuva:17d}. The amplitude for the fermion-fermion-dimer
recombination into two dimers equals to the three-cluster breakup amplitude
\begin{equation} \label{eq:U0}
\begin{split}  
 \langle \Phi_{3} |  \mct_{3 2} | \Phi_{2} \rangle 
= {}&  2  \langle \Phi_{3} | 
[(1- P_{34}) U_1 G_0 \, t \, G_0 \, \mcu_{12} \\ & {} +  
U_2 G_0 \,  t \, G_0 \, \mcu_{22} ]
| \phi_{2} \rangle ,
\end{split}
\end{equation}
where $ | \Phi_{3} \rangle$ abbreviates the three-cluster channel state and $  | \Phi_{2} \rangle $
the two-dimer channel state whose  Faddeev component is
$ |\phi_{2} \rangle = G_0 v | \Phi_{2} \rangle $
obeying also the Faddeev equation
$|\phi_{2} \rangle = G_0 t P_2 |\phi_{2} \rangle$.

The amplitude (\ref{eq:U0}) determines the three-cluster breakup and recombination observables.
The definition of the recombination rate $K_4$ follows from the number of recombination events
$K_4 \rho_d \rho_+ \rho_- $ per volume
and time, where $\rho_d$, $\rho_+$, and $\rho_-$ are
 densities of dimers and of fermions in spin-up and spin-down states, respectively.
Of special interest is the  fermion-fermion-dimer recombination rate $K_4^0$ at the threshold,
i.e., at vanishing three-cluster kinetic energy $E_3 = E + b_d \to 0$. This implies $k_y = k_z = 0$
and only $l_y = l_z = 0$ states contribute to this particular channel state
$ | \Phi_{3}^0 \rangle$. Furthermore,  all the discrete quantum numbers
take their minimum possible values of 0 or $\frac12$, i.e., the angular momentum
part of this state in the 2+2 basis reduces to a single component
$|\Phi_{\mcj=0}^0 \rangle = | (0  \{ [0 (\frac12\frac12)0]0\, [0 (\frac12\frac12)0] 0 \} 0) { 0 0} \rangle_2 $.
Since $\mcj$ is conserved and the dimer has zero spin,
the final two-dimer state $| \Phi_{2}^0 \rangle$ is restricted to have  exactly the same
angular momentum part $|\Phi_{\mcj=0}^0 \rangle$, while the relative momentum between the dimers is
$k_z = p_{dd}^0 = \sqrt{2mb_d}$.
Under these conditions the zero-energy  fermion-fermion-dimer recombination rate
is given by a single amplitude of the breakup operator in the partial-wave representation, i.e.,
\begin{equation} \label{eq:K40}
K_4^0 = {4 \pi^5} m p_{dd}^0
 | \langle \Phi_{3}^{0} |  \mct_{3 2} | \Phi_{2}^0 \rangle |^2 .
\end{equation}
It is important to note that, despite angular momentum limitations in the initial and final
channel states, the solution of the AGS equations (\ref{eq:U}) and the calculation
of the amplitude (\ref{eq:U0}) necessitates the inclusion of higher partial waves to achieve the convergence.
In fact, $l_y, l_z > 0$ waves contribute about 25\% to $K_4^0$  results in the next section.
A good convergence is achieved by including states with $l_y, l_z < 4$; it was found that those with
$l_y, l_z = 3 $ contribute about 0.2\%, implying that higher waves are negligible.

At finite $E_3$ a continuum of states contributes to the recombination, they have to be integrated
over either explicitly or implicitly, via the optical theorem. Below the four-particle
threshold the same integral determines also the dimer-dimer breakup cross section $\sigma_b$,
one of the standard observables in the scattering processes.
It is thus convenient to express the fermion-fermion-dimer recombination rate as
\begin{equation} \label{eq:K4e}
K_4 = \frac{8\sqrt{2} \pi  p_{dd}^2} {m^3\,E_3^2 } \, \sigma_b.
\end{equation}
The expressions (\ref{eq:K40}) and (\ref{eq:K4e}) take into account the identity of two fermions
(dimers) in the initial (final) states as well as the weight factors related to the spin averaging.

\section{Results \label{sec:res}}

A fermionic system with large scattering length is realized by neutrons, but
$a \approx -19$ fm  is negative and therefore no bound dineutron exists.
Nevertheless, for the present investigation I 
take a fictitious four-neutron system with slightly enhanced two-neutron force that
supports a bound two-particle state. Starting from the realistic CD Bonn potential \cite{machleidt:01a}
and fermion mass $m=938.9$ MeV,  enhancement factors from 1.35 to 1.105 lead to
scattering length $a$  from 9.1 to 176.1 fm, while the two-particle effective range
$r_e$ stays between  2.14 and 2.56 fm. Thus, within this relatively narrow variation of the
potential strength one can explore rather broad range of the scattering length as well
as of the ratio $r_e/a$ that quantifies the importance of finite-range effects.
Since the present study is devoted to the $S$-wave interacting four-fermion universality,
for the computational efficiency the potential in all higher two-fermion partial waves
with $l_x>0$ is assumed to be zero.

\begin{figure}[!]
\begin{center}
\includegraphics[scale=0.64]{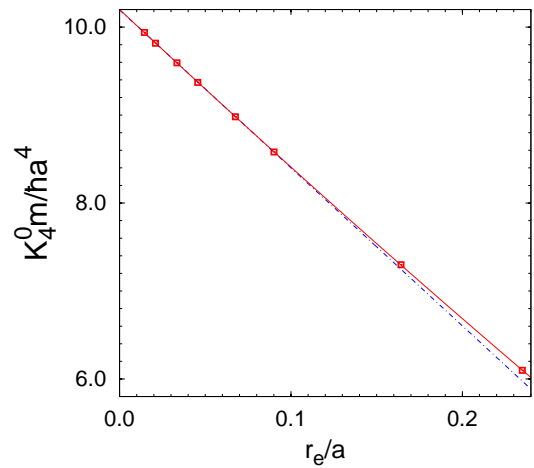}
\end{center}
\caption{\label{fig:k0} (Color online)
  Fermion-fermion-dimer recombination rate at zero energy
  in dimensionless form as a function of the finite-range parameter
  $r_e/a$. The symbols are the results of the calculation with the enhanced CD Bonn potential,
the solid curve represents Eq.~(\ref{eq:K4fit2}), while the dashed-dotted curve is a linear
approximation.
}
\end{figure}

The results of the above-described calculations for the fermion-fermion-dimer recombination
rate $K_4^0$ at the threshold are presented in Fig.~\ref{fig:k0}. Since the three-cluster
recombination rate scales as $a^4$, a dimensionless quantity
$K_4^0 m /\hbar a^4$ is build in order to demonstrate the universal behavior of this observable.
To a good accuracy, better than 1\% below $r_e/a = 0.1$,
 it appears to be linear in the finite-range correction parameter $r_e/a$,
  but the data points beyond this limit slightly violate this linear dependence. Nevertheless,
all data points  are well described including  finite-range corrections of a higher order, 
quadratic in $r_e/a$, resulting in an accurate analytical representation 
\begin{equation} \label{eq:K4fit2}
K_4^0 \approx \left[ 10.20 - 18.27 \frac{r_e}{a} + 3.47 \left(\frac{r_e}{a} \right)^2 \right] \, \frac{\hbar a^4}{m}.
\end{equation}
Obviously, in the considered regime the $(r_e/a)^2$ term yields only a minor correction.
Note that also the dimer-dimer scattering length and the effective range parameter
show a linear dependence on $r_e/a$ near the unitary limit, independently of the short-range details
of the employed potential \cite{deltuva:17d}. A tentative calculation using the separable interaction
model from Ref.~\cite{deltuva:17d} indicates this kind of independence also for $K_4^0 m /\hbar a^4$.
Thus, the relation (\ref{eq:K4fit2}) yields not only an accurate value for the ultracold
fermion-fermion-dimer recombination rate in the unitary limit but also reliably evaluates
the importance of the finite-range effects.

\begin{figure}[!]
\begin{center}
\includegraphics[scale=0.64]{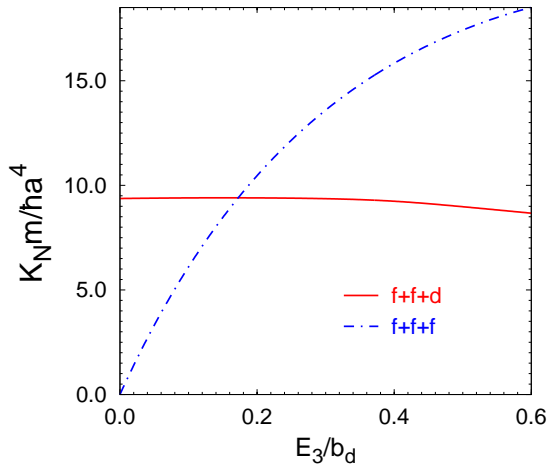}
\end{center}
\caption{\label{fig:ke} (Color online)
  Three-cluster recombination rates in three- and four-fermion systems,
displayed by the dashed-dotted  and solid curves, respectively, 
as functions  of the relative three-cluster kinetic  energy.
All quantities are shown in the dimensionless form and correspond to the
finite-range parameter $r_e/a = 0.0456$.
}
\end{figure}

Another question to be addressed is the energy dependence of the fermion-fermion-dimer recombination rate.
It is shown as a solid curve in Fig.~\ref{fig:ke} for $r_e/a = 0.0456$, since the shape of the energy dependence
is quite insensitive to the $r_e/a$ value provided it is below 0.1.
 The recombination rate depends weakly on the energy, indicating the absence of resonances as well as rather
insignificant contribution of $l_y,l_z > 0$ waves in the initial three-cluster channel.

It is interesting to compare the three-cluster recombination rates $K_N$ in  three- and four-fermion systems.
In the $N=3$ case the recombination rate $K_3$ vanishes at threshold as discussed in  Sec. I,
but the process is dominated by the total orbital momentum $\mcl=1$ state and therefore
$K_3$ increases with energy as shown by the dashed-dotted curve in  Fig.~\ref{fig:ke}.
At very low energy the increase of $K_3$ is linear as predicted in Ref.~\cite{esry:01a} by
analytical considerations, but starts to slow down for $ E_3 > 0.05 \, b_d$.
Only the $\mcl=1$ component is included in these results, since other components are suppressed even more
strongly, at least as $E_3^2$ \cite{esry:01a}, and were verified numerically  to be negligible in the low-energy regime.
Nevertheless, near  $ E_3 = 0.2 \, b_d$ the three-fermion recombination rate $K_3$ already
exceeds $K_4$. Note that  the  fermion-fermion-dimer recombination  in the $\mcl=1$ (equivalent to
$\mcj=1$) state is forbidden, since the final two-dimer state must be symmetric under the exchange of dimers,
which for zero-spin dimers is only possible for even $\mcl$ and $\mcj$ values.

The fact that the $\mcl=1$ process, quantified by $K_3$, already at relatively low energy exceeds
the  $\mcl=0$ dominated process, quantified by $K_4$,  may appear surprising, but has its explanation in the
partial suppression of the fermion-fermion-dimer recombination due to the four-fermion antisymmetry.
Although no antisymmetry restrictions apply to the dimer directly, 
the wave function must be antisymmetric under the exchange of free fermions with those inside the dimer,
thereby putting the  restrictions on the dimer indirectly. This is well reflected in the partial cancellation,
up to 90\%, of the 3+1 and 2+2 configuration contributions to the breakup or recombination amplitude (\ref{eq:U0}). 
This cancellation constitutes also a challenge in numerical calculations, that require more dense momentum
grids and smaller values for the imaginary part of the energy as compared to the
elastic scattering. %It is also interesting to note that
Another consequence of the antisymmetry restriction is that the 
fermion-fermion-dimer recombination rate significantly smaller than typical
values in other three-cluster systems \cite{braaten:rev}.

Finally,  there are important implications of the three-cluster recombination for the ultracold mixtures of
spin $\frac12$ fermions and dimers.
Let's consider an idealized system at zero temperature. If the system consists of unbound fermions, recombination
processes are suppressed and the system remains stable. However, even if very few dimers are added to this system, 
they initiate fermion-fermion-dimer recombination process producing more dimers, thereby increasing density
of dimers and further enhancing dimer production. Apart from this, the energy $b_d$ released in each
recombination event is initially taken by the two outgoing dimers, but after a series of elastic collisions
with fermions is transfered to fermions, that acquire small but finite energy as well. This implies that
also the three-fermion recombination becomes possible, contributing to further creation of dimers.
Which of these mechanisms is more important, depends on the initial fermion and dimer densities.
If they are of a comparable size, the four-particle process will dominate,
since $K_4 >> K_3$ for vanishing energy.
On the contrary, if the starting point is an ultracold  fermionic gas with a very small admixture of dimers $\rho_d^0$,
the fermion-fermion-dimer recombination would ``ignite'' the dimerization process, however,
the three-fermion recombination takes over once the fermions absorb
the released energy. A simple estimation is as follows:
At the time point when $\rho_d$ exceeds $\rho_d^0$ significantly,
the released energy is roughly $b_d \rho_d $ per unit volume,
thus, once dissipated between all particles through elastic two-cluster collisions,
it amounts to the kinetic energy per particle of roughly $b_d \rho_d/(\rho_+ + \rho_-)$.
This leads to the number of three-fermion recombination events per volume and time
$K_3 \rho_+\rho_-(\rho_+ + \rho_-) \approx  (\partial K_3/\partial E)|_{E=0} \, b_d \rho_+ \rho_- \rho_d$.
It is evident from  Fig.~\ref{fig:ke} that $(\partial K_3/\partial E)|_{E=0} \, b_d$ exceeds $K_4$,
thereby indicating the superiority of the three-fermion recombination under these conditions.

\section{Summary \label{sec:sum}}

The fermion-fermion-dimer recombination was studied using
exact scattering equations for the four-particle transition operators. 
They were solved numerically in the momentum-space partial-wave representation, employing the
complex-energy method with special integration weights for the treatment of kernel singularities.

In the systems of spin $\frac12$ fermions with resonant $S$-wave interactions
(and no dimers in other waves) the recombination
of three (or more) fermions at zero energy is forbidden by the antisymmetry requirement.
However, the fermion-fermion-dimer recombination is only partially suppressed,
having finite rate at the threshold. In the present work it was calculated for a rather broad range
of the two-fermion scattering length and to a good accuracy found to be linear in the effective range parameter $r_e/a$.
 This allowed for a reliable extrapolation to the unitary limit as well as the
evaluation of finite-range effects. The energy-dependence of the 
fermion-fermion-dimer recombination rate was shown to be weak, in contrast to the three-fermion recombination.
Their interplay was demonstrated to be important for the dimer production
in ultracold mixtures of fermions and dimers. 

One of the achievements of the present work, namely,  obtained universal result for the
zero energy fermion-fermion-dimer recombination rate, has important impact for the microscopic scattering
description in general.
 Being independent of the short-range interaction
details, it is well suitable for benchmarking  three-cluster breakup and recombination 
calculations in four-particle systems using different methods, that often have their preferred type
of the potential.

\vspace{1mm}

%The author acknowledges  support  by the Alexander von Humboldt Foundation under grant no. LTU-1185721-HFST-E.

%\clearpage
%%%%%%%%%%%%%%%%%%%%%%%%%%%%%%%%%%%%%%%%%%%%%%%%%%%%%%%%%%%%%%%%%%%%%%%%%%%%%

%\bibliographystyle{plbsty} 
%\bibliography{abbrev,pre80,80-89,90-99,200x,4N,ad,book,atomic,fermion,lowk,n34} \end{document}

\end{document}